\documentclass[%
aps,
pre,
amsmath,amssymb,
reprint,%
]{revtex4-1}
\usepackage{graphicx}
\usepackage{dcolumn}
\usepackage{bm}

\usepackage{amsmath}	
\begin{document}

\newcommand{\diff}[2]{\frac{d#1}{d#2}}
\newcommand{\pdiff}[2]{\frac{\partial #1}{\partial #2}}
\newcommand{\fdiff}[2]{\frac{\delta #1}{\delta #2}}
\newcommand{\bx}{\bm{x}}
\newcommand{\bq}{\bm{q}}
\newcommand{\br}{\bm{r}}
\newcommand{\bu}{\bm{u}}
\newcommand{\by}{\bm{y}}
\newcommand{\bY}{\bm{Y}}
\newcommand{\bF}{\bm{F}}
\newcommand{\new}{\nonumber\\}
\newcommand{\abs}[1]{\left|#1\right|}
\newcommand{\tr}{{\rm Tr}}
\newcommand{\HH}{{\mathcal H}}
\newcommand{\II}{{\mathcal I}}
\newcommand{\WW}{{\mathcal W}}
\newcommand{\OO}{{\mathcal O}}
\newcommand{\ave}[1]{\left\langle #1 \right\rangle}
\newcommand{\im}{{\rm Im}}
\newcommand{\re}{{\rm Re}}
\newcommand{\ke}{k_{\rm eff}}
\newcommand{\ipr}{{\rm IPR}}
\newcommand{\TT}{\tilde{T}}

\preprint{AIP/123-QED}

\title{Active Spherical Model}

\author{Harukuni Ikeda}
 \email{harukuni.ikeda@gakushuin.ac.jp}
\affiliation{ 
Department of Physics, Gakushuin University, 1-5-1 Mejiro, Toshima-ku, Tokyo 171-8588, Japan}

\date{\today}

\begin{abstract}
The spherical model is a popular solvable model and has been applied to
describe several critical phenomena such as the ferromagnetic
transition, Bose-Einstein condensation, spin-glass transition, glass
transition, jamming transition, and so on. Motivated by recent
developments of active matter, here we consider the spherical model
driven by the Ornstein–Uhlenbeck type self-propulsion force with
persistent time $\tau_p$. We show that the model exhibits the Ising
universality for finite $\tau_p$. On the contrary, the model exhibits
the random field Ising universality in the limit $\tau_p\to\infty$.
\end{abstract}

\maketitle

\section{Introduction}

In this work, we extend the spherical model to describe non-equilibrium
critical phenomena in the steady state. In particular, we have in mind
the so-called Motility-Induced Phase Separation (MIPS) of active matter:
the self-propelled particles spontaneously aggregate on increasing the
motility, eventually leading to a phase
separation~\cite{cates2015motility}. One of the fundamental questions is
to which universality class MIPS belongs. Extensive numerical
simulations showed that the MIPS has the Ising universality
class~\cite{partridge2019,maggi2021universality}, while a different
universality is also reported~\cite{siebert2018}. The field theoretical
studies support the Ising
universality~\cite{caballero2018bulk,cates2019active,maggi2022critical},
where the self-propulsion force on the $\phi^4$ field theory gives rise
to only irrelevant terms in the case of the bulk phase separation, while
a different universality appears in the case of the micro phase
separation~\cite{caballero2018bulk}. Here we study the effect of the
self-propulsion force on the Ising universality in a different way by
considering an exactly solvable model.

There are several solvable models to describe non-equilibrium critical
phenomena in the steady state, such as the celebrated Kuramoto
model~\cite{acebron2005kuramoto}, asymmetric simple exclusion
process~\cite{derrida1998exactly}, zero-range
process~\cite{evans2005nonequilibrium}, and so
on~\cite{tailleur2008,arnoulx2019}. However, most of those results are
obtained in the mean-field limit or one dimension, and it is not obvious
how to generalize the results to arbitrary spatial dimensions. The
spherical model is one of the few models that can be solved in any
dimension~\cite{berlin1952spherical} and several
settings~\cite{gunton1968condensation,kosterlitz1976,crisanti1992sphericalp,vojta1996,franz2016simplest,franz2017,casasola2021},
even out of
equilibrium~\cite{cugliandolo1993analytical,cugliandolo1995full,henkel2008non,berthier2013non,barbier2022generalised}.
Therefore, the spherical model would be the most promising candidate to
describe the non-equilibrium critical phenomena in steady states, such
as MIPS.

Here we consider the spherical model driven by a self-propulsion force
with persistent time $\tau_p$~\cite{ten2011brownian,szamel2014}.  We
show that for finite $\tau_p$, the model indeed has the Ising
universality, while in the limit $\tau_p\to\infty$, the model exhibits
the random field Ising universality.



The structure of the paper is as follows. In Sec.~\ref{113927_22Aug22},
we review some known results of the spherical model through the analysis
of the spherical Debye model (SDM), which is a simplified version of the
spherical model of a ferromagnet initially introduced by Berlin and
Kac~\cite{berlin1952spherical}. In particular, we argue that the
condensation transition of the model can be identified with the
underlying ferromagnetic transition~\cite{gunton1968condensation}. In
Sec.~\ref{114446_22Aug22}, we consider the SDM driven by an active noise
with the persistent time $\tau_p$. We investigate the model for
$\tau_p<\infty$. In Sec.~\ref{114623_22Aug22}, we investigate the model
in the limit $\tau_p\to\infty$. In Sec.~\ref{114656_22Aug22}, we briefly
discuss the effect of correlated noise. Finally, in
Sec.~\ref{114726_22Aug22}, we conclude the work.

\section{Spherical Debye model}
\label{113927_22Aug22}

Here we first review some known results for the spherical model through
the analysis of a simplified version of the spherical model originally
proposed by Berlin and Kac~\cite{berlin1952spherical}. We show that the
model exhibits the condensation transition at a critical temperature
$T_c$, and the transition has the same universality as the ferromagnetic
transition~\cite{berlin1952spherical,gunton1968condensation}.

Let we consider the following quadratic interaction potential:
\begin{align}
V_N = \frac{\bx\cdot W\cdot \bx}{2} + \frac{\mu}{2}(\bx\cdot\bx-N),\label{114924_11Aug22}
\end{align}
where $\bx=\{x_1,\dots, x_N\}$ denotes the state vector, and $W$ is a
$N\times N$ matrix representing the nearest neighbor interaction in a
$d$ dimensional lattice. $\mu$ denotes the Lagrange multiplier to impose
the following constraint:
\begin{align}
 \sum_{i=1}^N \ave{x_i^2} = N,\label{104848_19Aug22}
\end{align}
where the braket denotes the thermal average by the Maxwell-Boltzmann
distribution at temperature $T$. To analyze the model, one 
can expand $V_N$ by the normal modes:
\begin{align}
 V_N = \sum_{i=1}^N \frac{\omega_i^2+\mu}{2} u_i^2 -\frac{N\mu}{2},\label{110655_19Aug22}
\end{align}
where $\omega_i$ denotes the frequency of the $i$-th mode. We will order
$\omega_i$ as 
\begin{align}
 \omega_1 < \omega_2 < \cdots < \omega_N.
\end{align}
Since, an orthogonal transformation does not change the inner product,
the spherical constraint is written as
\begin{align}
 \sum_{i=1}^N \ave{u_i^2} = N.\label{130057_19Aug22}
\end{align}
The precise value of $\omega_i$ depends on the details of $W$, but the
vibrational properties of a $d$ dimensional lattice would eventually be
dominated by the phonon modes on a large enough scale. So we assume that
the distribution of $\omega_i$,
$D(\omega)=N^{-1}\sum_{i=1}^N\delta(\omega-\omega_i)$, is given by the
Debye density of states~\cite{kittel2018introduction}:
\begin{align}
D(\omega)=
\begin{cases}
d\omega_D^{-d}\omega^{d-1} & \omega\in [0,\omega_D],\\
 0 & {\rm otherwise},
\end{cases}
 \label{120601_11Aug22}
\end{align}
where $\omega_D$ denotes the Debye frequency.  $D(\omega)$ is normalized
so that $\int_0^{\omega_D}D(\omega)d\omega = 1$. To simplify the notation,
hereafter we set $\omega_D=1$. We call the model defined by
Eqs.~(\ref{110655_19Aug22}), (\ref{130057_19Aug22}), and
(\ref{120601_11Aug22}) as the spherical Debye model (SDM).

In this section, we study the model in equilibrium at temperature
$T$. From the equipartition theorem~\cite{greiner2012thermodynamics}, we
get
\begin{align}
\ave{u_i^2} = \frac{k_B T}{\omega_i^2+\mu},\label{151706_8Aug22}
\end{align}
where $k_B$ denotes the Boltzmann constant. To simplify the notation,
hereafter we set $k_B=1$. Since $\ave{u_i^2}\leq N$, $\mu$ should
satisfy the following condition
\begin{align}
\mu\geq -\min_i\omega_i^2 = 0.\label{153029_8Aug22}
\end{align}
The value of $\mu$ is to be determined by the spherical constraint
\begin{align}
1 = \frac{1}{N}\sum_{i=1}^N \frac{T}{\omega_i^2+\mu},\label{151627_8Aug22}
\end{align}
In the limit $N\to\infty$, we expect that the summation can be replaced
with an integral:
\begin{align}
&1=T F(\mu), &F(\mu) =\int d\omega  \frac{D(\omega)}{\omega^2+\mu}.
\label{155303_8Aug22}
\end{align}
$F(\mu)$ is a decreasing function of $\mu$
and is maximal in the limit $\mu\to +0$:
\begin{align}
\lim_{\mu\to +0}F(\mu)=  
\begin{cases}
+\infty & (d\leq 2)\\
 1/T_c & (d>2).
\end{cases}, 
\end{align}
where
\begin{align}
T_c = \frac{d-2}{2}.
\end{align}
When $d>2$ and $T<T_c$,
\begin{align}
 TF(\mu)\leq TF(0)<1,
\end{align}
implying that $TF(\mu)=1$ has no solution. This is the signature of the
condensation to the lowest eigenmode
$\omega_1$~\cite{gunton1968condensation,crisanti2019}. Below $T_c$, we
should separate the first and other terms in Eq.~(\ref{151627_8Aug22})
to replace the summation with an integral, as in the case of the
Bose-Einstein
condensation~\cite{greiner2012thermodynamics,crisanti2019}:
\begin{align}
\frac{1}{N}\sum_{i=1}^N \frac{T}{\omega_i^2+\mu}
& = \frac{\ave{u_1^2}}{N} + \frac{1}{N}\sum_{i=2}^N \frac{T}{\omega_i^2+\mu}\new 
& = \frac{\ave{u_1}^2}{N} + TF(\mu).\label{161103_21Aug22}
\end{align}
Substituting it back into Eq.~(\ref{151627_8Aug22}), we get for $T<T_c$
\begin{align}
\frac{\ave{u_1}^2}{N} = 1- TF(0) = 1- \frac{T}{T_c}.\label{161129_21Aug22}
\end{align}
In the case of the spherical model of a
ferromagnet~\cite{berlin1952spherical}, the condensation transition is
identified with the ferromagnetic phase transition, and
Eq.~(\ref{161129_21Aug22}) corresponds to the square of the
magnetization $\ave{u_1^2}/N\sim m^2$, which leads to the well-known
scaling behavior ${m\sim
(1-T/T_c)^{1/2}}$~\cite{berlin1952spherical,gunton1968condensation,crisanti2019}.

The detailed analysis of Eq.~(\ref{155303_8Aug22}) reveals that on
approaching $T_c$ from above, $\mu$ behaves as follows (see
Appendix.~\ref{123702_19Aug22}):
\begin{align}
 \mu \sim
 \begin{cases}
 (T-T_c)^{\frac{2}{d-2}} & d\in (2, 4),\\
 (T-T_c)^1 & d>4.
 \end{cases}\label{123351_19Aug22}
\end{align}
Below $T_c$, $\mu$ is written as 
\begin{align}
 \mu = \frac{1}{\ave{u_1}^2} = \frac{1}{N}\frac{T_c}{T_c-T},
\end{align}
which vanishes in the thermodynamic limit $N\to\infty$. The Lagrange
multiplier $\mu$ of the spherical model plays a similar role as the
chemical potential of the ideal Bose gas, in that it fixes the mean
value of an extensive quantity. Indeed, both quantities have the same
critical
exponent~\cite{gunton1968condensation,greiner2012thermodynamics,crisanti2019}.
Since $\ave{u_1^2}=1/\mu$, we get
\begin{align}
 \ave{u_1^2} \sim
 \begin{cases}
 (T-T_c)^{-\frac{2}{d-2}} & d\in (2, 4),\\
 (T-T_c)^{-1} & d>4.
 \end{cases},
\end{align}
which can be identified with the susceptibility $\chi \sim m^2$ of the
spherical ferromagnetic model for
$T>T_c$~\cite{gunton1968condensation,crisanti2019}.
Using the
equipartition theorem, the mean-value of the interaction potential
Eq.~(\ref{110655_19Aug22}) is calculated as follows
\begin{align}
u = \frac{\ave{V_N}}{N} = T -\mu
 = \begin{cases}
    T-\mu & (T> T_c)\\
    T & (T\leq T_c)
   \end{cases}.
\end{align}
The specific heat for $d\in (2,4)$ is 
\begin{align}
 C =\diff{u}{T} =
\begin{cases}
 (T-T_c)^{\frac{4-d}{d-2}} & (T>T_c)\\
 {\rm const}  & ( T<T_c)
\end{cases},
\end{align}
and for $d>4$, the critical exponent is zero, i.e., $C$ changes
discontinuously at $T=T_c$. The result is again consistent with the
spherical ferromagnetic model~\cite{berlin1952spherical}. One can also
argue the scaling of the correlation length by assuming the linear
relation $\omega_i = cq_i$ between the frequency $\omega_i$ and wave
number $q_i$~\cite{gunton1968condensation,nishimori2010elements}.

In summary, the SDM, which consists of $N$ non-interacting oscillators
Eq.~(\ref{110655_19Aug22}), has the same universality as the spherical
model of a ferromagnet. It is worth mentioning that the phase behavior
and the critical exponent of $\mu$ can be deduced from only the
information of the second moment of $u_i$ in the steady state
$\ave{u_i^2}$. This allows us to draw the phase diagram even for a
non-equilibrium version of the model where the steady-state distribution
is in general not known. Furthermore, the critical exponent of the
Lagrange multiplier $\mu$, which controls the other critical exponents,
is also calculated from $\ave{u_i^2}$. This allows us to discuss the
lower and upper critical dimensions, and the universality class.

\section{Active spherical model}
\label{114446_22Aug22}

Now we consider a non-equilibrium model.  We consider the following
equation~\cite{ten2011brownian}:
\begin{align}
\diff{u_i(t)}{t} &=
-\pdiff{V_N}{u_i(t)} + f_i(t)\new
 &= -(\mu+\omega_i^2)u_i(t) + f_i(t),\label{041056_21Aug22}
\end{align}
where $f_i(t)$ denotes the self-propulsion force.
The time evolution of $f_i(t)$ is given by the Ornstein-Uhlenbeck process~\cite{szamel2014}:
\begin{align}
 \tau_p \dot{f}_i(t) = -f_i(t) + \sqrt{2T}\eta_i(t),\label{121229_26Aug22}
\end{align}
where $\tau_p$ denotes the persistent time,
$T$ denotes the strength of the noise, 
and  $\eta_i(t)$ denotes the white noise
satisfying the following condition 
\begin{align}
&\ave{\eta_i(t)} =0,
&\ave{\eta_i(t)\eta_j(t')} = \delta_{ij}\delta(t-t').
\end{align}
Eq.~(\ref{121229_26Aug22}) can be directly integrated as
follows:
\begin{align}
 f_i(t) = \frac{\sqrt{2T}}{\tau_p}\int_{-\infty}^t dt' e^{-(t-t')/\tau_p}\xi_i(t').
\end{align}
Then, we get
\begin{align}
 \ave{f_i(t)f_j(t')} &= \frac{2T}{\tau_p^2}\int_{-\infty}^t dt_1 \int_{-\infty}^{t'}dt_2 e^{-(2t-t_1-t_2)/\tau_p}\new 
 &\times \delta_{ij}\delta(t_1-t_2)\new
 &= \delta_{ij}\frac{T}{\tau_p}e^{-\abs{t-t'}/\tau_p}.\label{122532_26Aug22}
\end{align}
In the limit $\tau_p\to 0$, $\ave{f_i(t)f_j(t')}\to
2T\delta_{ij}\delta(t-t')$, implying that $f_i(t)$ reduces to the
thermal white noise at temperature
$T$~\cite{zwanzig2001nonequilibrium}. Now we assume that the system is
in the steady state at $t=-\infty$ so that the Lagrange multiplier $\mu$
does not depend on time. Then, one can easily integrate
Eq.~(\ref{041056_21Aug22}) as follows:
\begin{align}
u_i(t) = \int_{-\infty}^t dt' e^{-(\mu+\omega_i^2)(t-t')}f_i(t').
\end{align}
With a similar calculation as Eq.~(\ref{122532_26Aug22}), the second
moment of $u_i(t)$ in the steady state is calculated as
follows~\cite{szamel2014}:
\begin{align}
\ave{u_i(t)^2} = \frac{T}{(\mu+\omega_i^2)\left[1+\tau_p(\mu+\omega_i^2)\right]}.\label{123714_21Aug22}
\end{align}
This implies that the equipartition theorem Eq.~(\ref{151706_8Aug22})
does not hold for $\tau_p>0$, as a consequence of the violation of the
detailed balance~\cite{szamel2014}. Some readers may wonder how $f_i(t)$
behaves in real space. In fact, the properties of $f_i(t)$ do not depend
on the choice of the coordinate system, since a transformation by an
orthogonal matrix $O$ preserves the second moment of $f_i(t)$:
\begin{align}
\ave{(Of(t))_i(Of(t'))_j} &= 
 \sum_{nm}O_{in}O_{jm}\ave{f_n(t)f_m(t')}\new 
&= \frac{T}{\tau_p} \delta_{ij}e^{-\abs{t_1-t_2}/\tau_p}.
\end{align}

\begin{figure}[t]
\begin{center}
\includegraphics[width=9cm]{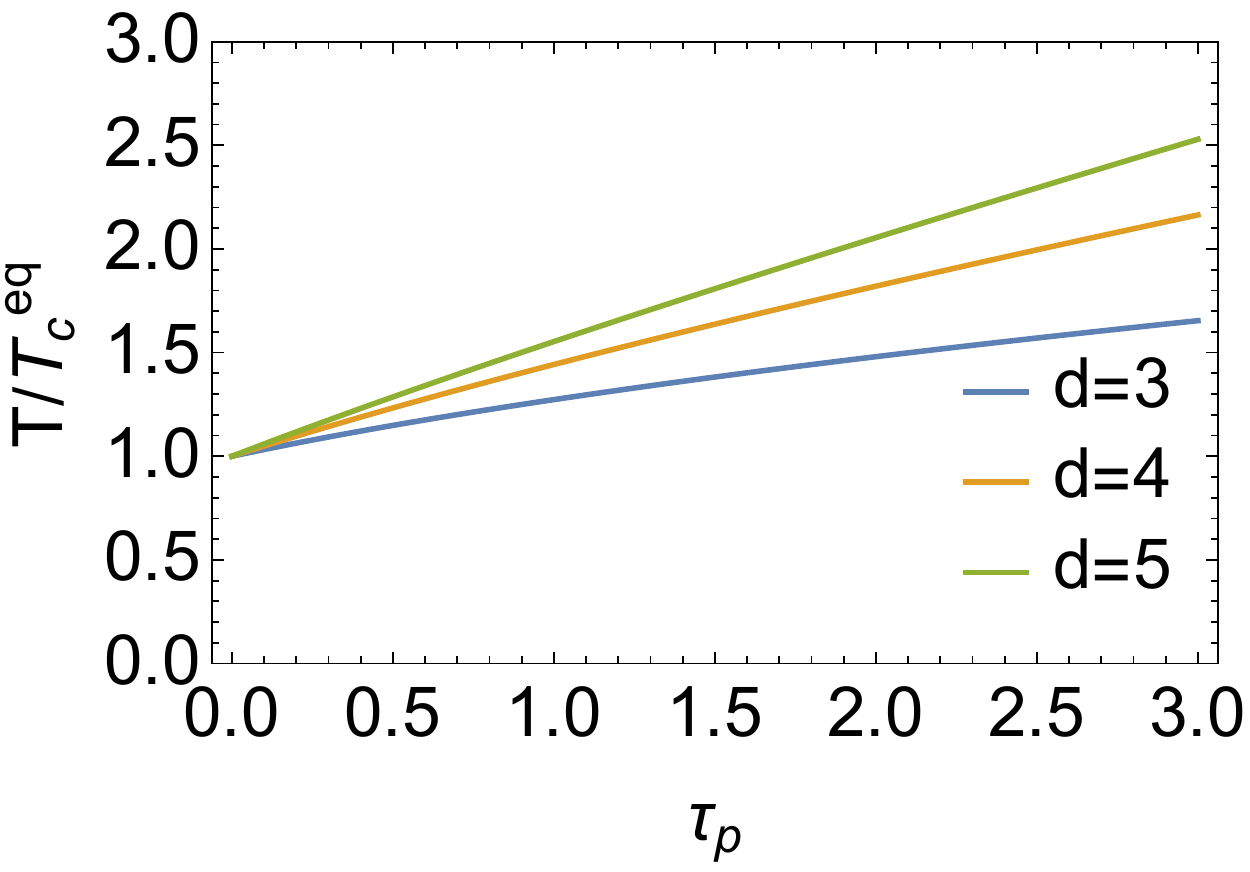} \caption{Phase diagram of the
active spherical model. The vertical axis is rescaled by the critical
temperature in equilibrium $T_c^{\rm eq}\equiv T_c(\tau_p=0)$.  The
solid line denotes the transition line.}  \label{154537_21Aug22}
\end{center}
\end{figure}

The value of $\mu$ is to be determined by the spherical constraint: $1 =
\frac{1}{N}\sum_{i=1}^N \ave{u_i^2}$. For $T>T_c$, the summation can be
replaced with an integral:
\begin{align}
&1=T F(\mu),\new 
&F(\mu)=\int d\omega D(\omega)\frac{1}{(\mu+\omega^2)\left[1+\tau_p(\mu+\omega^2)\right]}.\label{172036_21Aug22}
\end{align}
For finite $\tau_p$ and for $d>2$, $F(\mu)$ converges to a finite value,
implying that the condensation transition occurs at the critical
temperature
\begin{align}
T_c = \frac{1}{F(0)}.
\end{align}
As shown in Fig.~\ref{154537_21Aug22}, $T_c$ increases on increasing
$\tau_p$, suggesting that the motility facilitates the phase transition,
as in the case of the MIPS~\cite{cates2015motility}.
Below $T_c$, the condensation to the first mode occurs as in the case of
the equilibrium model, see Eq.~(\ref{161129_21Aug22}). The critical
behavior for $\mu\ll 1$ is governed by the small $\omega$ behavior of the integrand 
in Eq.~(\ref{172036_21Aug22}):
\begin{align}
 \frac{D(\omega)}{(\mu+\omega^2)\left[1+\tau_p(\mu+\omega^2)\right]}\sim
 \frac{D(\omega)}{(\mu+\omega^2)},
\end{align}
which is the same as that of the original model
Eq.~(\ref{155303_8Aug22}). Therefore, the critical exponent is unchanged
from the original one.

\section{Extreme active matter}
\label{114623_22Aug22} Recently, the extreme active matter, which
corresponds to the limit $\tau_p\to \infty$, has attracted much
attention~\cite{mandal2020extreme,caprini2020,morse2021direct}. So we
here consider the corresponding limit. For this purpose, we introduce
a scaled variable:
\begin{align}
\TT \equiv \frac{T}{\tau_p}.
\end{align}
Then, in the limit $\tau_p\to \infty$, Eq.~(\ref{123714_21Aug22})
reduces to
\begin{align}
\ave{u_i^2} \to \frac{\TT}{(\mu+\omega_i)^2}.\label{191158_6Sep22}
\end{align}
Repeating the same arguments as in the previous sections, we obtain
the self-consistent equation for $\mu$:
\begin{align}
&1 = \TT F(\mu),\new
&F(\mu) = \int d\omega \frac{D(\omega)}{(\omega^2+\mu)^2}.\label{162957_21Aug22}
\end{align}
In the limit $\mu\to 0$, we get
\begin{align}
\lim_{\mu\to0}F(\mu) =
 \begin{cases}
  \infty  & (d\leq 4)\\
  d/(d-4) & (d>4)
 \end{cases},
\end{align}
implying that the lower critical dimension is $d_u=4$.  Below $T<T_c$,
the condensation to the first mode occurs:
\begin{align}
\frac{\ave{u_1^2}}{N} = 1-\frac{T}{T_c},
\end{align}
where
\begin{align}
T_c = F(0)^{-1} = \frac{d-4}{d}.
\end{align}
The detailed analysis of Eq.~(\ref{162957_21Aug22}) leads to
\begin{align}
 \mu &=
 \begin{cases}
  (T-T_c)^{\frac{2}{d-4}} & d\in (4,6)\\
  (T-T_c)^{1} & d>6,
  \end{cases},\\
 \ave{u_1^2} &=
 \begin{cases}
  (T-T_c)^{-\frac{4}{d-4}} & d\in (4,6)\\
  (T-T_c)^{-2} & d>6,
 \end{cases}
\end{align}
implying that the upper critical dimension is $d_u=6$, see
Appendix.~\ref{123702_19Aug22}. The above results are consistent with
the spherical model with random field~\cite{vojta1996random}. This would
be a reasonable result because in the large $\tau_p$ limit, the
self-propulsion force is permanently frozen and may be identified with a
random field, see also Appendix~\ref{191507_6Sep22}.

\section{Correlated noise} 
\label{114656_22Aug22} The analysis in the previous sections revealed
that the phase transition does not occur for $d\leq d_u=2$ when
$\tau_p<\infty$ and for $d\leq d_u=4$ when $\tau_p=\infty$. However,
there is some numerical evidence of the phase transitions in
$d=2$~\cite{siebert2018,partridge2019}. A possible ingredient of the
phase transition in low $d$ is a long-range correlation of $f_i$, which
originates from the hydrodynamic interaction~\cite{landau2013fluid},
elastic interaction~\cite{landau1986theory}, or something
else~\cite{szamel2021long,kuroda2022anomalous}. So, here we briefly
discuss the effects of the correlated noise. For this purpose, we modify
Eq.~(\ref{122532_26Aug22}) as follows:
\begin{align}
\ave{f_i(t)f_j(t')} = \delta_{ij}\frac{T_i}{\tau_p}e^{-\abs{t-t'}/\tau_p},
\end{align}
where $T_i$ depends on the frequency $\omega_i$ (or equivalently wave
number $q_i=c^{-1}\omega_i$). The long-range correlation would suppress
the fluctuation of $f_i$ on a large scale, {\it i.e.}, small
$q_i=c^{-1}\omega_i$. To express this effect, we assume a power-law
function:
\begin{align}
T_i = T\omega_i^p.\label{172437_21Aug22}
\end{align}
Now the self-consistent equation for $\mu$ Eq.~(\ref{172036_21Aug22}) is
modified as
\begin{align}
&1 = TF(\mu),\new 
&F(\mu)= \int_0^1 d\omega 
\frac{\omega^{d+p-1}}{(\mu+\omega^2)\left[1+\tau_p(\mu+\omega^2)\right]}.
\end{align}
From the above expression, one can see that the correlated noise
Eq.~(\ref{172437_21Aug22}) effectively increases the spatial dimension
from $d$ to $d+p$.
Therefore, the upper and lower critical dimensions are
\begin{align}
d_l =
 \begin{cases}
  2-p & (\tau_p<\infty)\\
  4-p & (\tau_p=\infty)
 \end{cases},\\
 d_u  =
 \begin{cases}
  4-p & (\tau_p<\infty)\\
  6-p & (\tau_p=\infty)
 \end{cases}.
\end{align}
Therefore, the $\omega_i$ (or $q_i$) dependence of the noise
Eq.~(\ref{172437_21Aug22}) indeed can cause the phase transition below
the lower critical dimension. However in reality, the noise should be
determined self-consistently as a consequence of the complex interaction
between the system and environment~\cite{sagu2007}. Further studies for
this problem would be beneficial.

\section{Summary and discussions}
\label{114726_22Aug22}

In summary, we investigated the effects of the active noise on the
spherical model. For this purpose, we first introduced and analyzed the
spherical Debye model, which is a simplified version of the spherical
model for a ferromagnet. The model is so simple that one can discuss the
phase behavior solely from the information of the second-order moment at
the steady state, even without complete knowledge of the steady state
distribution. The simplicity also allows us to introduce an active
noise: the noise produced by the Ornstein–Uhlenbeck process with the
persistent time $\tau_p$. We found that for a finite value of $\tau_p$,
the model has the same universality as the original model, while
in the limit $\tau_p\to\infty$, the universality reduces to that of
the random field Ising model.

In this work, we only investigated the static properties of the model.
Of course, what is more interesting is the dynamic behaviors such as
aging dynamics~\cite{cugliandolo1995full,henkel2022quantum}, effective
temperature~\cite{cugliandolo1993analytical,berthier2013non}, entropy
production~\cite{cates2020stealth,matteo2022}, and so on. The current
model may allow us to derive a full dynamical solution as done for
$p$-spin spherical
models~\cite{cugliandolo1993analytical,cugliandolo1995full}. Further
studies on this problem would be beneficial.

\acknowledgments

This project has received JSPS KAKENHI Grant Numbers 21K20355.

\appendix

\section{Critical exponent}
\label{123702_19Aug22}

To determine $\mu$, one should solve the following self-consistent equation:
\begin{align}
1 = TF(\mu)\equiv TA\int_0^1 d\omega  \frac{\omega^{d-1}}{(\omega^2+\mu)^n},\label{122931_12Aug22}
\end{align}
where $A$ is a constant, $n=1$ for $\tau_p<\infty$ and $n=2$ for
$\tau_p=\infty$.
If $d-2n>0$, $F(0)$ is finite, implying that there exists the condensation
transition. So, the lower critical dimension is
\begin{align}
 d_u = 2n.
\end{align}
When $d>2n+2$, $F(\mu)$ can be expanded as 
\begin{align}
\frac{1}{T} &= F(0) + \mu F'(0) + \cdots\new 
 &= \frac{1}{T_c} + \mu F'(0) + \cdots,
\end{align}
leading to
\begin{align}
\mu \sim (T-T_c)^{1}.
\end{align}
On the contrary, if $d\in (2n,2n+2)$, $F'(\mu)$ for small $\mu$ 
behaves as 
\begin{align}
 F'(\mu) \sim \mu^{\frac{d-(2n+2)}{2}},\label{120735_13Aug22}
\end{align}
implying 
\begin{align}
 F(\mu) -F(0) =\int_0^{\mu}d\mu' F'(\mu') \sim \mu^{\frac{d-2n}{2}},
\end{align}
leading to
\begin{align}
 \frac{1}{T} = F(\mu) = \frac{1}{T_c}-B\mu^{\frac{d-2n}{2}},
\end{align}
where $B$ is a constant. Therefore, the scaling of $\mu$ for $\mu\ll 1$ is 
\begin{align}
\mu \sim (T-T_c)^{\frac{2}{d-2n}}.
\end{align}
A logaithmic correction may appear when
$d=2n+2$~\cite{nishimori2010elements}.  The above results imply that the
upper critical dimension is
\begin{align}
 d_u = 2n+2.
\end{align}

\section{Spherical model with random field}
\label{191507_6Sep22}
We here consider the spherical model with a random
field~\cite{vojta1996random,ikeda2022sat}:
\begin{align}
 V_N = \sum_{i=1}^N \frac{\lambda_i+\mu}{2}u_i^2 -\frac{N\mu}{2}
 + \sum_{i=1}^N h_i u_i,
\end{align}
where $h_i$ is an i.i.d random variable of zero mean and variance $\Delta$.
In equilibrium at temperature $T$, we get~\cite{ikeda2022sat}
\begin{align}
 \overline{\ave{u_i^2}} = \frac{T}{\omega_i^2+\mu} + \frac{\Delta}{(\omega_i^2+\mu)^2},\label{191007_6Sep22}
\end{align}
where the overline denotes the average for $h_i$, and
$\Delta=\overline{h_i^2}$.  The Lagrange multiplier $\mu$ is to be
determined by the spherical constraint:
\begin{align}
1 = \frac{1}{N}\sum_{i=1}^N
\left[\frac{T}{\omega_i^2+\mu} + \frac{\Delta}{(\omega_i^2+\mu)^2}\right].\label{174319_16Aug22}
\end{align}
As mentioned in the main text, the scaling for $\mu\ll 1$ is determined
by the small $\omega_i$ behavior of $\overline{\ave{u_i^2}}$, where the
second term in Eq.~(\ref{191007_6Sep22}) gives a dominant contribution:
\begin{align}
\overline{\ave{u_i^2}} \approx\frac{\Delta}{(\omega_i^2+\mu)^2}.
\end{align}
This agrees with the active spherical model with $\tau_p\to\infty$,
Eq.~(\ref{191158_6Sep22}). Therefore, the active spherical model with
$\tau_p\to \infty$ can be identified with the spherical model with
random field of variance $\Delta=\tilde{T}$.

\bibliography{reference}

\begin{thebibliography}{45}%
\makeatletter
\providecommand \@ifxundefined [1]{%
 \@ifx{#1\undefined}
}%
\providecommand \@ifnum [1]{%
 \ifnum #1\expandafter \@firstoftwo
 \else \expandafter \@secondoftwo
 \fi
}%
\providecommand \@ifx [1]{%
 \ifx #1\expandafter \@firstoftwo
 \else \expandafter \@secondoftwo
 \fi
}%
\providecommand \natexlab [1]{#1}%
\providecommand \enquote  [1]{``#1''}%
\providecommand \bibnamefont  [1]{#1}%
\providecommand \bibfnamefont [1]{#1}%
\providecommand \citenamefont [1]{#1}%
\providecommand \href@noop [0]{\@secondoftwo}%
\providecommand \href [0]{\begingroup \@sanitize@url \@href}%
\providecommand \@href[1]{\@@startlink{#1}\@@href}%
\providecommand \@@href[1]{\endgroup#1\@@endlink}%
\providecommand \@sanitize@url [0]{\catcode `\\12\catcode `\$12\catcode
  `\&12\catcode `\#12\catcode `\^12\catcode `\_12\catcode `\%12\relax}%
\providecommand \@@startlink[1]{}%
\providecommand \@@endlink[0]{}%
\providecommand \url  [0]{\begingroup\@sanitize@url \@url }%
\providecommand \@url [1]{\endgroup\@href {#1}{\urlprefix }}%
\providecommand \urlprefix  [0]{URL }%
\providecommand \Eprint [0]{\href }%
\providecommand \doibase [0]{http://dx.doi.org/}%
\providecommand \selectlanguage [0]{\@gobble}%
\providecommand \bibinfo  [0]{\@secondoftwo}%
\providecommand \bibfield  [0]{\@secondoftwo}%
\providecommand \translation [1]{[#1]}%
\providecommand \BibitemOpen [0]{}%
\providecommand \bibitemStop [0]{}%
\providecommand \bibitemNoStop [0]{.\EOS\space}%
\providecommand \EOS [0]{\spacefactor3000\relax}%
\providecommand \BibitemShut  [1]{\csname bibitem#1\endcsname}%
\let\auto@bib@innerbib\@empty
\bibitem [{\citenamefont {Cates}\ and\ \citenamefont
  {Tailleur}(2015)}]{cates2015motility}%
  \BibitemOpen
  \bibfield  {author} {\bibinfo {author} {\bibfnamefont {M.~E.}\ \bibnamefont
  {Cates}}\ and\ \bibinfo {author} {\bibfnamefont {J.}~\bibnamefont
  {Tailleur}},\ }\href@noop {} {\bibfield  {journal} {\bibinfo  {journal}
  {Annu. Rev. Condens. Matter Phys.}\ }\textbf {\bibinfo {volume} {6}},\
  \bibinfo {pages} {219} (\bibinfo {year} {2015})}\BibitemShut {NoStop}%
\bibitem [{\citenamefont {Partridge}\ and\ \citenamefont
  {Lee}(2019)}]{partridge2019}%
  \BibitemOpen
  \bibfield  {author} {\bibinfo {author} {\bibfnamefont {B.}~\bibnamefont
  {Partridge}}\ and\ \bibinfo {author} {\bibfnamefont {C.~F.}\ \bibnamefont
  {Lee}},\ }\href {\doibase 10.1103/PhysRevLett.123.068002} {\bibfield
  {journal} {\bibinfo  {journal} {Phys. Rev. Lett.}\ }\textbf {\bibinfo
  {volume} {123}},\ \bibinfo {pages} {068002} (\bibinfo {year}
  {2019})}\BibitemShut {NoStop}%
\bibitem [{\citenamefont {Maggi}\ \emph {et~al.}(2021)\citenamefont {Maggi},
  \citenamefont {Paoluzzi}, \citenamefont {Crisanti}, \citenamefont
  {Zaccarelli},\ and\ \citenamefont {Gnan}}]{maggi2021universality}%
  \BibitemOpen
  \bibfield  {author} {\bibinfo {author} {\bibfnamefont {C.}~\bibnamefont
  {Maggi}}, \bibinfo {author} {\bibfnamefont {M.}~\bibnamefont {Paoluzzi}},
  \bibinfo {author} {\bibfnamefont {A.}~\bibnamefont {Crisanti}}, \bibinfo
  {author} {\bibfnamefont {E.}~\bibnamefont {Zaccarelli}}, \ and\ \bibinfo
  {author} {\bibfnamefont {N.}~\bibnamefont {Gnan}},\ }\href@noop {} {\bibfield
   {journal} {\bibinfo  {journal} {Soft Matter}\ }\textbf {\bibinfo {volume}
  {17}},\ \bibinfo {pages} {3807} (\bibinfo {year} {2021})}\BibitemShut
  {NoStop}%
\bibitem [{\citenamefont {Siebert}\ \emph {et~al.}(2018)\citenamefont
  {Siebert}, \citenamefont {Dittrich}, \citenamefont {Schmid}, \citenamefont
  {Binder}, \citenamefont {Speck},\ and\ \citenamefont {Virnau}}]{siebert2018}%
  \BibitemOpen
  \bibfield  {author} {\bibinfo {author} {\bibfnamefont {J.~T.}\ \bibnamefont
  {Siebert}}, \bibinfo {author} {\bibfnamefont {F.}~\bibnamefont {Dittrich}},
  \bibinfo {author} {\bibfnamefont {F.}~\bibnamefont {Schmid}}, \bibinfo
  {author} {\bibfnamefont {K.}~\bibnamefont {Binder}}, \bibinfo {author}
  {\bibfnamefont {T.}~\bibnamefont {Speck}}, \ and\ \bibinfo {author}
  {\bibfnamefont {P.}~\bibnamefont {Virnau}},\ }\href {\doibase
  10.1103/PhysRevE.98.030601} {\bibfield  {journal} {\bibinfo  {journal} {Phys.
  Rev. E}\ }\textbf {\bibinfo {volume} {98}},\ \bibinfo {pages} {030601}
  (\bibinfo {year} {2018})}\BibitemShut {NoStop}%
\bibitem [{\citenamefont {Caballero}\ \emph {et~al.}(2018)\citenamefont
  {Caballero}, \citenamefont {Nardini},\ and\ \citenamefont
  {Cates}}]{caballero2018bulk}%
  \BibitemOpen
  \bibfield  {author} {\bibinfo {author} {\bibfnamefont {F.}~\bibnamefont
  {Caballero}}, \bibinfo {author} {\bibfnamefont {C.}~\bibnamefont {Nardini}},
  \ and\ \bibinfo {author} {\bibfnamefont {M.~E.}\ \bibnamefont {Cates}},\
  }\href@noop {} {\bibfield  {journal} {\bibinfo  {journal} {Journal of
  Statistical Mechanics: Theory and Experiment}\ }\textbf {\bibinfo {volume}
  {2018}},\ \bibinfo {pages} {123208} (\bibinfo {year} {2018})}\BibitemShut
  {NoStop}%
\bibitem [{\citenamefont {Cates}(2019)}]{cates2019active}%
  \BibitemOpen
  \bibfield  {author} {\bibinfo {author} {\bibfnamefont {M.~E.}\ \bibnamefont
  {Cates}},\ }\href@noop {} {\bibfield  {journal} {\bibinfo  {journal} {arXiv
  preprint arXiv:1904.01330}\ } (\bibinfo {year} {2019})}\BibitemShut {NoStop}%
\bibitem [{\citenamefont {Maggi}\ \emph {et~al.}(2022)\citenamefont {Maggi},
  \citenamefont {Gnan}, \citenamefont {Paoluzzi}, \citenamefont {Zaccarelli},\
  and\ \citenamefont {Crisanti}}]{maggi2022critical}%
  \BibitemOpen
  \bibfield  {author} {\bibinfo {author} {\bibfnamefont {C.}~\bibnamefont
  {Maggi}}, \bibinfo {author} {\bibfnamefont {N.}~\bibnamefont {Gnan}},
  \bibinfo {author} {\bibfnamefont {M.}~\bibnamefont {Paoluzzi}}, \bibinfo
  {author} {\bibfnamefont {E.}~\bibnamefont {Zaccarelli}}, \ and\ \bibinfo
  {author} {\bibfnamefont {A.}~\bibnamefont {Crisanti}},\ }\href@noop {}
  {\bibfield  {journal} {\bibinfo  {journal} {Communications Physics}\ }\textbf
  {\bibinfo {volume} {5}},\ \bibinfo {pages} {1} (\bibinfo {year}
  {2022})}\BibitemShut {NoStop}%
\bibitem [{\citenamefont {Acebr{\'o}n}\ \emph {et~al.}(2005)\citenamefont
  {Acebr{\'o}n}, \citenamefont {Bonilla}, \citenamefont {Vicente},
  \citenamefont {Ritort},\ and\ \citenamefont {Spigler}}]{acebron2005kuramoto}%
  \BibitemOpen
  \bibfield  {author} {\bibinfo {author} {\bibfnamefont {J.~A.}\ \bibnamefont
  {Acebr{\'o}n}}, \bibinfo {author} {\bibfnamefont {L.~L.}\ \bibnamefont
  {Bonilla}}, \bibinfo {author} {\bibfnamefont {C.~J.~P.}\ \bibnamefont
  {Vicente}}, \bibinfo {author} {\bibfnamefont {F.}~\bibnamefont {Ritort}}, \
  and\ \bibinfo {author} {\bibfnamefont {R.}~\bibnamefont {Spigler}},\
  }\href@noop {} {\bibfield  {journal} {\bibinfo  {journal} {Reviews of modern
  physics}\ }\textbf {\bibinfo {volume} {77}},\ \bibinfo {pages} {137}
  (\bibinfo {year} {2005})}\BibitemShut {NoStop}%
\bibitem [{\citenamefont {Derrida}(1998)}]{derrida1998exactly}%
  \BibitemOpen
  \bibfield  {author} {\bibinfo {author} {\bibfnamefont {B.}~\bibnamefont
  {Derrida}},\ }\href@noop {} {\bibfield  {journal} {\bibinfo  {journal}
  {Physics Reports}\ }\textbf {\bibinfo {volume} {301}},\ \bibinfo {pages} {65}
  (\bibinfo {year} {1998})}\BibitemShut {NoStop}%
\bibitem [{\citenamefont {Evans}\ and\ \citenamefont
  {Hanney}(2005)}]{evans2005nonequilibrium}%
  \BibitemOpen
  \bibfield  {author} {\bibinfo {author} {\bibfnamefont {M.~R.}\ \bibnamefont
  {Evans}}\ and\ \bibinfo {author} {\bibfnamefont {T.}~\bibnamefont {Hanney}},\
  }\href@noop {} {\bibfield  {journal} {\bibinfo  {journal} {Journal of Physics
  A: Mathematical and General}\ }\textbf {\bibinfo {volume} {38}},\ \bibinfo
  {pages} {R195} (\bibinfo {year} {2005})}\BibitemShut {NoStop}%
\bibitem [{\citenamefont {Tailleur}\ and\ \citenamefont
  {Cates}(2008)}]{tailleur2008}%
  \BibitemOpen
  \bibfield  {author} {\bibinfo {author} {\bibfnamefont {J.}~\bibnamefont
  {Tailleur}}\ and\ \bibinfo {author} {\bibfnamefont {M.~E.}\ \bibnamefont
  {Cates}},\ }\href {\doibase 10.1103/PhysRevLett.100.218103} {\bibfield
  {journal} {\bibinfo  {journal} {Phys. Rev. Lett.}\ }\textbf {\bibinfo
  {volume} {100}},\ \bibinfo {pages} {218103} (\bibinfo {year}
  {2008})}\BibitemShut {NoStop}%
\bibitem [{\citenamefont {Arnoulx~de Pirey}\ \emph {et~al.}(2019)\citenamefont
  {Arnoulx~de Pirey}, \citenamefont {Lozano},\ and\ \citenamefont {van
  Wijland}}]{arnoulx2019}%
  \BibitemOpen
  \bibfield  {author} {\bibinfo {author} {\bibfnamefont {T.}~\bibnamefont
  {Arnoulx~de Pirey}}, \bibinfo {author} {\bibfnamefont {G.}~\bibnamefont
  {Lozano}}, \ and\ \bibinfo {author} {\bibfnamefont {F.}~\bibnamefont {van
  Wijland}},\ }\href {\doibase 10.1103/PhysRevLett.123.260602} {\bibfield
  {journal} {\bibinfo  {journal} {Phys. Rev. Lett.}\ }\textbf {\bibinfo
  {volume} {123}},\ \bibinfo {pages} {260602} (\bibinfo {year}
  {2019})}\BibitemShut {NoStop}%
\bibitem [{\citenamefont {Berlin}\ and\ \citenamefont
  {Kac}(1952)}]{berlin1952spherical}%
  \BibitemOpen
  \bibfield  {author} {\bibinfo {author} {\bibfnamefont {T.~H.}\ \bibnamefont
  {Berlin}}\ and\ \bibinfo {author} {\bibfnamefont {M.}~\bibnamefont {Kac}},\
  }\href@noop {} {\bibfield  {journal} {\bibinfo  {journal} {Physical Review}\
  }\textbf {\bibinfo {volume} {86}},\ \bibinfo {pages} {821} (\bibinfo {year}
  {1952})}\BibitemShut {NoStop}%
\bibitem [{\citenamefont {Gunton}\ and\ \citenamefont
  {Buckingham}(1968)}]{gunton1968condensation}%
  \BibitemOpen
  \bibfield  {author} {\bibinfo {author} {\bibfnamefont {J.}~\bibnamefont
  {Gunton}}\ and\ \bibinfo {author} {\bibfnamefont {M.}~\bibnamefont
  {Buckingham}},\ }\href@noop {} {\bibfield  {journal} {\bibinfo  {journal}
  {Physical Review}\ }\textbf {\bibinfo {volume} {166}},\ \bibinfo {pages}
  {152} (\bibinfo {year} {1968})}\BibitemShut {NoStop}%
\bibitem [{\citenamefont {Kosterlitz}\ \emph {et~al.}(1976)\citenamefont
  {Kosterlitz}, \citenamefont {Thouless},\ and\ \citenamefont
  {Jones}}]{kosterlitz1976}%
  \BibitemOpen
  \bibfield  {author} {\bibinfo {author} {\bibfnamefont {J.~M.}\ \bibnamefont
  {Kosterlitz}}, \bibinfo {author} {\bibfnamefont {D.~J.}\ \bibnamefont
  {Thouless}}, \ and\ \bibinfo {author} {\bibfnamefont {R.~C.}\ \bibnamefont
  {Jones}},\ }\href {\doibase 10.1103/PhysRevLett.36.1217} {\bibfield
  {journal} {\bibinfo  {journal} {Phys. Rev. Lett.}\ }\textbf {\bibinfo
  {volume} {36}},\ \bibinfo {pages} {1217} (\bibinfo {year}
  {1976})}\BibitemShut {NoStop}%
\bibitem [{\citenamefont {Crisanti}\ and\ \citenamefont
  {Sommers}(1992)}]{crisanti1992sphericalp}%
  \BibitemOpen
  \bibfield  {author} {\bibinfo {author} {\bibfnamefont {A.}~\bibnamefont
  {Crisanti}}\ and\ \bibinfo {author} {\bibfnamefont {H.-J.}\ \bibnamefont
  {Sommers}},\ }\href@noop {} {\bibfield  {journal} {\bibinfo  {journal}
  {Zeitschrift f{\"u}r Physik B Condensed Matter}\ }\textbf {\bibinfo {volume}
  {87}},\ \bibinfo {pages} {341} (\bibinfo {year} {1992})}\BibitemShut
  {NoStop}%
\bibitem [{\citenamefont {Vojta}(1996)}]{vojta1996}%
  \BibitemOpen
  \bibfield  {author} {\bibinfo {author} {\bibfnamefont {T.}~\bibnamefont
  {Vojta}},\ }\href {\doibase 10.1103/PhysRevB.53.710} {\bibfield  {journal}
  {\bibinfo  {journal} {Phys. Rev. B}\ }\textbf {\bibinfo {volume} {53}},\
  \bibinfo {pages} {710} (\bibinfo {year} {1996})}\BibitemShut {NoStop}%
\bibitem [{\citenamefont {Franz}\ and\ \citenamefont
  {Parisi}(2016)}]{franz2016simplest}%
  \BibitemOpen
  \bibfield  {author} {\bibinfo {author} {\bibfnamefont {S.}~\bibnamefont
  {Franz}}\ and\ \bibinfo {author} {\bibfnamefont {G.}~\bibnamefont {Parisi}},\
  }\href@noop {} {\bibfield  {journal} {\bibinfo  {journal} {Journal of Physics
  A: Mathematical and Theoretical}\ }\textbf {\bibinfo {volume} {49}},\
  \bibinfo {pages} {145001} (\bibinfo {year} {2016})}\BibitemShut {NoStop}%
\bibitem [{\citenamefont {Franz}\ \emph {et~al.}(2017)\citenamefont {Franz},
  \citenamefont {Parisi}, \citenamefont {Sevelev}, \citenamefont {Urbani},\
  and\ \citenamefont {Zamponi}}]{franz2017}%
  \BibitemOpen
  \bibfield  {author} {\bibinfo {author} {\bibfnamefont {S.}~\bibnamefont
  {Franz}}, \bibinfo {author} {\bibfnamefont {G.}~\bibnamefont {Parisi}},
  \bibinfo {author} {\bibfnamefont {M.}~\bibnamefont {Sevelev}}, \bibinfo
  {author} {\bibfnamefont {P.}~\bibnamefont {Urbani}}, \ and\ \bibinfo {author}
  {\bibfnamefont {F.}~\bibnamefont {Zamponi}},\ }\href {\doibase
  10.21468/SciPostPhys.2.3.019} {\bibfield  {journal} {\bibinfo  {journal}
  {SciPost Phys.}\ }\textbf {\bibinfo {volume} {2}},\ \bibinfo {pages} {019}
  (\bibinfo {year} {2017})}\BibitemShut {NoStop}%
\bibitem [{\citenamefont {Casasola}\ \emph {et~al.}(2021)\citenamefont
  {Casasola}, \citenamefont {Hernaski}, \citenamefont {Gomes},\ and\
  \citenamefont {Bienzobaz}}]{casasola2021}%
  \BibitemOpen
  \bibfield  {author} {\bibinfo {author} {\bibfnamefont {H.}~\bibnamefont
  {Casasola}}, \bibinfo {author} {\bibfnamefont {C.~A.}\ \bibnamefont
  {Hernaski}}, \bibinfo {author} {\bibfnamefont {P.~R.~S.}\ \bibnamefont
  {Gomes}}, \ and\ \bibinfo {author} {\bibfnamefont {P.~F.}\ \bibnamefont
  {Bienzobaz}},\ }\href {\doibase 10.1103/PhysRevE.104.034131} {\bibfield
  {journal} {\bibinfo  {journal} {Phys. Rev. E}\ }\textbf {\bibinfo {volume}
  {104}},\ \bibinfo {pages} {034131} (\bibinfo {year} {2021})}\BibitemShut
  {NoStop}%
\bibitem [{\citenamefont {Cugliandolo}\ and\ \citenamefont
  {Kurchan}(1993)}]{cugliandolo1993analytical}%
  \BibitemOpen
  \bibfield  {author} {\bibinfo {author} {\bibfnamefont {L.~F.}\ \bibnamefont
  {Cugliandolo}}\ and\ \bibinfo {author} {\bibfnamefont {J.}~\bibnamefont
  {Kurchan}},\ }\href@noop {} {\bibfield  {journal} {\bibinfo  {journal}
  {Physical Review Letters}\ }\textbf {\bibinfo {volume} {71}},\ \bibinfo
  {pages} {173} (\bibinfo {year} {1993})}\BibitemShut {NoStop}%
\bibitem [{\citenamefont {Cugliandolo}\ and\ \citenamefont
  {Dean}(1995)}]{cugliandolo1995full}%
  \BibitemOpen
  \bibfield  {author} {\bibinfo {author} {\bibfnamefont {L.~F.}\ \bibnamefont
  {Cugliandolo}}\ and\ \bibinfo {author} {\bibfnamefont {D.~S.}\ \bibnamefont
  {Dean}},\ }\href@noop {} {\bibfield  {journal} {\bibinfo  {journal} {Journal
  of Physics A: Mathematical and General}\ }\textbf {\bibinfo {volume} {28}},\
  \bibinfo {pages} {4213} (\bibinfo {year} {1995})}\BibitemShut {NoStop}%
\bibitem [{\citenamefont {Henkel}\ \emph {et~al.}(2008)\citenamefont {Henkel},
  \citenamefont {Hinrichsen}, \citenamefont {L{\"u}beck},\ and\ \citenamefont
  {Pleimling}}]{henkel2008non}%
  \BibitemOpen
  \bibfield  {author} {\bibinfo {author} {\bibfnamefont {M.}~\bibnamefont
  {Henkel}}, \bibinfo {author} {\bibfnamefont {H.}~\bibnamefont {Hinrichsen}},
  \bibinfo {author} {\bibfnamefont {S.}~\bibnamefont {L{\"u}beck}}, \ and\
  \bibinfo {author} {\bibfnamefont {M.}~\bibnamefont {Pleimling}},\ }\href@noop
  {} {\emph {\bibinfo {title} {Non-equilibrium phase transitions}}},\
  Vol.~\bibinfo {volume} {1}\ (\bibinfo  {publisher} {Springer},\ \bibinfo
  {year} {2008})\BibitemShut {NoStop}%
\bibitem [{\citenamefont {Berthier}\ and\ \citenamefont
  {Kurchan}(2013)}]{berthier2013non}%
  \BibitemOpen
  \bibfield  {author} {\bibinfo {author} {\bibfnamefont {L.}~\bibnamefont
  {Berthier}}\ and\ \bibinfo {author} {\bibfnamefont {J.}~\bibnamefont
  {Kurchan}},\ }\href@noop {} {\bibfield  {journal} {\bibinfo  {journal}
  {Nature Physics}\ }\textbf {\bibinfo {volume} {9}},\ \bibinfo {pages} {310}
  (\bibinfo {year} {2013})}\BibitemShut {NoStop}%
\bibitem [{\citenamefont {Barbier}\ \emph {et~al.}(2022)\citenamefont
  {Barbier}, \citenamefont {Cugliandolo}, \citenamefont {Lozano},\ and\
  \citenamefont {Nessi}}]{barbier2022generalised}%
  \BibitemOpen
  \bibfield  {author} {\bibinfo {author} {\bibfnamefont {D.}~\bibnamefont
  {Barbier}}, \bibinfo {author} {\bibfnamefont {L.~F.}\ \bibnamefont
  {Cugliandolo}}, \bibinfo {author} {\bibfnamefont {G.~S.}\ \bibnamefont
  {Lozano}}, \ and\ \bibinfo {author} {\bibfnamefont {N.}~\bibnamefont
  {Nessi}},\ }\href@noop {} {\bibfield  {journal} {\bibinfo  {journal} {arXiv
  preprint arXiv:2204.03081}\ } (\bibinfo {year} {2022})}\BibitemShut {NoStop}%
\bibitem [{\citenamefont {ten Hagen}\ \emph {et~al.}(2011)\citenamefont {ten
  Hagen}, \citenamefont {van Teeffelen},\ and\ \citenamefont
  {L{\"o}wen}}]{ten2011brownian}%
  \BibitemOpen
  \bibfield  {author} {\bibinfo {author} {\bibfnamefont {B.}~\bibnamefont {ten
  Hagen}}, \bibinfo {author} {\bibfnamefont {S.}~\bibnamefont {van Teeffelen}},
  \ and\ \bibinfo {author} {\bibfnamefont {H.}~\bibnamefont {L{\"o}wen}},\
  }\href@noop {} {\bibfield  {journal} {\bibinfo  {journal} {Journal of
  Physics: Condensed Matter}\ }\textbf {\bibinfo {volume} {23}},\ \bibinfo
  {pages} {194119} (\bibinfo {year} {2011})}\BibitemShut {NoStop}%
\bibitem [{\citenamefont {Szamel}(2014)}]{szamel2014}%
  \BibitemOpen
  \bibfield  {author} {\bibinfo {author} {\bibfnamefont {G.}~\bibnamefont
  {Szamel}},\ }\href {\doibase 10.1103/PhysRevE.90.012111} {\bibfield
  {journal} {\bibinfo  {journal} {Phys. Rev. E}\ }\textbf {\bibinfo {volume}
  {90}},\ \bibinfo {pages} {012111} (\bibinfo {year} {2014})}\BibitemShut
  {NoStop}%
\bibitem [{\citenamefont {Kittel}\ and\ \citenamefont
  {McEuen}(2018)}]{kittel2018introduction}%
  \BibitemOpen
  \bibfield  {author} {\bibinfo {author} {\bibfnamefont {C.}~\bibnamefont
  {Kittel}}\ and\ \bibinfo {author} {\bibfnamefont {P.}~\bibnamefont
  {McEuen}},\ }\href@noop {} {\emph {\bibinfo {title} {Introduction to solid
  state physics}}}\ (\bibinfo  {publisher} {John Wiley \& Sons},\ \bibinfo
  {year} {2018})\BibitemShut {NoStop}%
\bibitem [{\citenamefont {Greiner}\ \emph {et~al.}(2012)\citenamefont
  {Greiner}, \citenamefont {Neise},\ and\ \citenamefont
  {St{\"o}cker}}]{greiner2012thermodynamics}%
  \BibitemOpen
  \bibfield  {author} {\bibinfo {author} {\bibfnamefont {W.}~\bibnamefont
  {Greiner}}, \bibinfo {author} {\bibfnamefont {L.}~\bibnamefont {Neise}}, \
  and\ \bibinfo {author} {\bibfnamefont {H.}~\bibnamefont {St{\"o}cker}},\
  }\href@noop {} {\emph {\bibinfo {title} {Thermodynamics and statistical
  mechanics}}}\ (\bibinfo  {publisher} {Springer Science \& Business Media},\
  \bibinfo {year} {2012})\BibitemShut {NoStop}%
\bibitem [{\citenamefont {Crisanti}\ \emph {et~al.}(2019)\citenamefont
  {Crisanti}, \citenamefont {Sarracino},\ and\ \citenamefont
  {Zannetti}}]{crisanti2019}%
  \BibitemOpen
  \bibfield  {author} {\bibinfo {author} {\bibfnamefont {A.}~\bibnamefont
  {Crisanti}}, \bibinfo {author} {\bibfnamefont {A.}~\bibnamefont {Sarracino}},
  \ and\ \bibinfo {author} {\bibfnamefont {M.}~\bibnamefont {Zannetti}},\
  }\href {\doibase 10.1103/PhysRevResearch.1.023022} {\bibfield  {journal}
  {\bibinfo  {journal} {Phys. Rev. Research}\ }\textbf {\bibinfo {volume}
  {1}},\ \bibinfo {pages} {023022} (\bibinfo {year} {2019})}\BibitemShut
  {NoStop}%
\bibitem [{\citenamefont {Nishimori}\ and\ \citenamefont
  {Ortiz}(2010)}]{nishimori2010elements}%
  \BibitemOpen
  \bibfield  {author} {\bibinfo {author} {\bibfnamefont {H.}~\bibnamefont
  {Nishimori}}\ and\ \bibinfo {author} {\bibfnamefont {G.}~\bibnamefont
  {Ortiz}},\ }\href@noop {} {\emph {\bibinfo {title} {Elements of phase
  transitions and critical phenomena}}}\ (\bibinfo  {publisher} {Oup Oxford},\
  \bibinfo {year} {2010})\BibitemShut {NoStop}%
\bibitem [{\citenamefont {Zwanzig}(2001)}]{zwanzig2001nonequilibrium}%
  \BibitemOpen
  \bibfield  {author} {\bibinfo {author} {\bibfnamefont {R.}~\bibnamefont
  {Zwanzig}},\ }\href@noop {} {\emph {\bibinfo {title} {Nonequilibrium
  statistical mechanics}}}\ (\bibinfo  {publisher} {Oxford university press},\
  \bibinfo {year} {2001})\BibitemShut {NoStop}%
\bibitem [{\citenamefont {Mandal}\ \emph {et~al.}(2020)\citenamefont {Mandal},
  \citenamefont {Bhuyan}, \citenamefont {Chaudhuri}, \citenamefont {Dasgupta},\
  and\ \citenamefont {Rao}}]{mandal2020extreme}%
  \BibitemOpen
  \bibfield  {author} {\bibinfo {author} {\bibfnamefont {R.}~\bibnamefont
  {Mandal}}, \bibinfo {author} {\bibfnamefont {P.~J.}\ \bibnamefont {Bhuyan}},
  \bibinfo {author} {\bibfnamefont {P.}~\bibnamefont {Chaudhuri}}, \bibinfo
  {author} {\bibfnamefont {C.}~\bibnamefont {Dasgupta}}, \ and\ \bibinfo
  {author} {\bibfnamefont {M.}~\bibnamefont {Rao}},\ }\href@noop {} {\bibfield
  {journal} {\bibinfo  {journal} {Nature communications}\ }\textbf {\bibinfo
  {volume} {11}},\ \bibinfo {pages} {1} (\bibinfo {year} {2020})}\BibitemShut
  {NoStop}%
\bibitem [{\citenamefont {Caprini}\ \emph {et~al.}(2020)\citenamefont
  {Caprini}, \citenamefont {Marconi}, \citenamefont {Maggi}, \citenamefont
  {Paoluzzi},\ and\ \citenamefont {Puglisi}}]{caprini2020}%
  \BibitemOpen
  \bibfield  {author} {\bibinfo {author} {\bibfnamefont {L.}~\bibnamefont
  {Caprini}}, \bibinfo {author} {\bibfnamefont {U.~M.~B.}\ \bibnamefont
  {Marconi}}, \bibinfo {author} {\bibfnamefont {C.}~\bibnamefont {Maggi}},
  \bibinfo {author} {\bibfnamefont {M.}~\bibnamefont {Paoluzzi}}, \ and\
  \bibinfo {author} {\bibfnamefont {A.}~\bibnamefont {Puglisi}},\ }\href
  {\doibase 10.1103/PhysRevResearch.2.023321} {\bibfield  {journal} {\bibinfo
  {journal} {Phys. Rev. Research}\ }\textbf {\bibinfo {volume} {2}},\ \bibinfo
  {pages} {023321} (\bibinfo {year} {2020})}\BibitemShut {NoStop}%
\bibitem [{\citenamefont {Morse}\ \emph {et~al.}(2021)\citenamefont {Morse},
  \citenamefont {Roy}, \citenamefont {Agoritsas}, \citenamefont {Stanifer},
  \citenamefont {Corwin},\ and\ \citenamefont {Manning}}]{morse2021direct}%
  \BibitemOpen
  \bibfield  {author} {\bibinfo {author} {\bibfnamefont {P.~K.}\ \bibnamefont
  {Morse}}, \bibinfo {author} {\bibfnamefont {S.}~\bibnamefont {Roy}}, \bibinfo
  {author} {\bibfnamefont {E.}~\bibnamefont {Agoritsas}}, \bibinfo {author}
  {\bibfnamefont {E.}~\bibnamefont {Stanifer}}, \bibinfo {author}
  {\bibfnamefont {E.~I.}\ \bibnamefont {Corwin}}, \ and\ \bibinfo {author}
  {\bibfnamefont {M.~L.}\ \bibnamefont {Manning}},\ }\href@noop {} {\bibfield
  {journal} {\bibinfo  {journal} {Proceedings of the National Academy of
  Sciences}\ }\textbf {\bibinfo {volume} {118}},\ \bibinfo {pages}
  {e2019909118} (\bibinfo {year} {2021})}\BibitemShut {NoStop}%
\bibitem [{\citenamefont {Vojta}\ and\ \citenamefont
  {Schreiber}(1996)}]{vojta1996random}%
  \BibitemOpen
  \bibfield  {author} {\bibinfo {author} {\bibfnamefont {T.}~\bibnamefont
  {Vojta}}\ and\ \bibinfo {author} {\bibfnamefont {M.}~\bibnamefont
  {Schreiber}},\ }\href {\doibase 10.1103/PhysRevB.53.8211} {\bibfield
  {journal} {\bibinfo  {journal} {Phys. Rev. B}\ }\textbf {\bibinfo {volume}
  {53}},\ \bibinfo {pages} {8211} (\bibinfo {year} {1996})}\BibitemShut
  {NoStop}%
\bibitem [{\citenamefont {Landau}\ and\ \citenamefont
  {Lifshitz}(2013)}]{landau2013fluid}%
  \BibitemOpen
  \bibfield  {author} {\bibinfo {author} {\bibfnamefont {L.~D.}\ \bibnamefont
  {Landau}}\ and\ \bibinfo {author} {\bibfnamefont {E.~M.}\ \bibnamefont
  {Lifshitz}},\ }\href@noop {} {\emph {\bibinfo {title} {Fluid Mechanics:
  Landau and Lifshitz: Course of Theoretical Physics, Volume 6}}},\
  Vol.~\bibinfo {volume} {6}\ (\bibinfo  {publisher} {Elsevier},\ \bibinfo
  {year} {2013})\BibitemShut {NoStop}%
\bibitem [{\citenamefont {Landau}\ \emph {et~al.}(1986)\citenamefont {Landau},
  \citenamefont {Lif{\v{s}}ic}, \citenamefont {Lifshitz}, \citenamefont
  {Kosevich},\ and\ \citenamefont {Pitaevskii}}]{landau1986theory}%
  \BibitemOpen
  \bibfield  {author} {\bibinfo {author} {\bibfnamefont {L.~D.}\ \bibnamefont
  {Landau}}, \bibinfo {author} {\bibfnamefont {E.~M.}\ \bibnamefont
  {Lif{\v{s}}ic}}, \bibinfo {author} {\bibfnamefont {E.~M.}\ \bibnamefont
  {Lifshitz}}, \bibinfo {author} {\bibfnamefont {A.~M.}\ \bibnamefont
  {Kosevich}}, \ and\ \bibinfo {author} {\bibfnamefont {L.~P.}\ \bibnamefont
  {Pitaevskii}},\ }\href@noop {} {\emph {\bibinfo {title} {Theory of
  elasticity: volume 7}}},\ Vol.~\bibinfo {volume} {7}\ (\bibinfo  {publisher}
  {Elsevier},\ \bibinfo {year} {1986})\BibitemShut {NoStop}%
\bibitem [{\citenamefont {Szamel}\ and\ \citenamefont
  {Flenner}(2021)}]{szamel2021long}%
  \BibitemOpen
  \bibfield  {author} {\bibinfo {author} {\bibfnamefont {G.}~\bibnamefont
  {Szamel}}\ and\ \bibinfo {author} {\bibfnamefont {E.}~\bibnamefont
  {Flenner}},\ }\href@noop {} {\bibfield  {journal} {\bibinfo  {journal}
  {Europhysics Letters}\ }\textbf {\bibinfo {volume} {133}},\ \bibinfo {pages}
  {60002} (\bibinfo {year} {2021})}\BibitemShut {NoStop}%
\bibitem [{\citenamefont {Kuroda}\ \emph {et~al.}(2022)\citenamefont {Kuroda},
  \citenamefont {Matsuyama}, \citenamefont {Kawasaki},\ and\ \citenamefont
  {Miyazaki}}]{kuroda2022anomalous}%
  \BibitemOpen
  \bibfield  {author} {\bibinfo {author} {\bibfnamefont {Y.}~\bibnamefont
  {Kuroda}}, \bibinfo {author} {\bibfnamefont {H.}~\bibnamefont {Matsuyama}},
  \bibinfo {author} {\bibfnamefont {T.}~\bibnamefont {Kawasaki}}, \ and\
  \bibinfo {author} {\bibfnamefont {K.}~\bibnamefont {Miyazaki}},\ }\href@noop
  {} {\bibfield  {journal} {\bibinfo  {journal} {arXiv preprint
  arXiv:2202.04436}\ } (\bibinfo {year} {2022})}\BibitemShut {NoStop}%
\bibitem [{\citenamefont {Sagu\'es}\ \emph {et~al.}(2007)\citenamefont
  {Sagu\'es}, \citenamefont {Sancho},\ and\ \citenamefont
  {Garc\'{\i}a-Ojalvo}}]{sagu2007}%
  \BibitemOpen
  \bibfield  {author} {\bibinfo {author} {\bibfnamefont {F.}~\bibnamefont
  {Sagu\'es}}, \bibinfo {author} {\bibfnamefont {J.~M.}\ \bibnamefont
  {Sancho}}, \ and\ \bibinfo {author} {\bibfnamefont {J.}~\bibnamefont
  {Garc\'{\i}a-Ojalvo}},\ }\href {\doibase 10.1103/RevModPhys.79.829}
  {\bibfield  {journal} {\bibinfo  {journal} {Rev. Mod. Phys.}\ }\textbf
  {\bibinfo {volume} {79}},\ \bibinfo {pages} {829} (\bibinfo {year}
  {2007})}\BibitemShut {NoStop}%
\bibitem [{\citenamefont {Henkel}(2022)}]{henkel2022quantum}%
  \BibitemOpen
  \bibfield  {author} {\bibinfo {author} {\bibfnamefont {M.}~\bibnamefont
  {Henkel}},\ }\href@noop {} {\bibfield  {journal} {\bibinfo  {journal} {arXiv
  preprint arXiv:2201.06448}\ } (\bibinfo {year} {2022})}\BibitemShut {NoStop}%
\bibitem [{\citenamefont {Caballero}\ and\ \citenamefont
  {Cates}(2020)}]{cates2020stealth}%
  \BibitemOpen
  \bibfield  {author} {\bibinfo {author} {\bibfnamefont {F.}~\bibnamefont
  {Caballero}}\ and\ \bibinfo {author} {\bibfnamefont {M.~E.}\ \bibnamefont
  {Cates}},\ }\href {\doibase 10.1103/PhysRevLett.124.240604} {\bibfield
  {journal} {\bibinfo  {journal} {Phys. Rev. Lett.}\ }\textbf {\bibinfo
  {volume} {124}},\ \bibinfo {pages} {240604} (\bibinfo {year}
  {2020})}\BibitemShut {NoStop}%
\bibitem [{\citenamefont {Paoluzzi}(2022)}]{matteo2022}%
  \BibitemOpen
  \bibfield  {author} {\bibinfo {author} {\bibfnamefont {M.}~\bibnamefont
  {Paoluzzi}},\ }\href {\doibase 10.1103/PhysRevE.105.044139} {\bibfield
  {journal} {\bibinfo  {journal} {Phys. Rev. E}\ }\textbf {\bibinfo {volume}
  {105}},\ \bibinfo {pages} {044139} (\bibinfo {year} {2022})}\BibitemShut
  {NoStop}%
\bibitem [{\citenamefont {Ikeda}(2022)}]{ikeda2022sat}%
  \BibitemOpen
  \bibfield  {author} {\bibinfo {author} {\bibfnamefont {H.}~\bibnamefont
  {Ikeda}},\ }\href@noop {} {\bibfield  {journal} {\bibinfo  {journal} {arXiv
  preprint arXiv:2208.08162}\ } (\bibinfo {year} {2022})}\BibitemShut {NoStop}%
\end{thebibliography}%
\end{document}